\documentclass[a4paper, twocolumn]{article}

\usepackage[dvips]{graphicx}
\usepackage{amsmath, amssymb, url, cite, balance}

\newcommand{\mv}[3]{#1_#2^{(#3)}}

\title{A simple view of the heavy-tailed sales distributions and application to the box-office grosses of U.S. movies}
\author{Ken Yamamoto}
\newcommand{\institute}[1]{\date{\small #1}}
\institute{Department of Physics, Faculty of Science and Engineering, Chuo University, Kasuga 1, Bunkyo, Tokyo 112--8551, Japan}

\begin{document}
\twocolumn[\begin{@twocolumnfalse}
\maketitle

\begin{abstract}
This letter treats of the power-law distribution of the sales of items.
We propose a simple stochastic model which expresses a selling process of an item.
This model produces a stationary power-law distribution, whose power-law exponent is analytically derived.
Next we compare the model with an actual data set of movie income.
We focus on the return on investment (ROI), defined as the gross income divided by the production budget.
We confirm that the power-law exponent of ROI distribution can be estimated from the ratios of income between two adjoining weeks, as predicted by the model analysis.
Moreover, exponential decay of weekly income is observed both in the model and actual income.
Therefore, the proposed model is simple enough, but it can quantitatively describe the power-law sales distribution.
\end{abstract}
\vspace*{5mm}
\end{@twocolumnfalse}
]

The heavy-tailed distributions, including the power-law and log-normal distributions, have been found in various social phenomena~\cite{Buchanan, Kobayashi}.
For example, the stock volatility~\cite{Gabaix}, the human mobility~\cite{Rhee}, and the population of cities \cite{Zanette} follow power-law distributions,
and the citation of physics papers \cite{Redner}, the polling score of election \cite{Fortunato} and the population of villages \cite{Sasaki} follow log-normal distributions.
A remarkable implication of the heavy-tailed distribution is that it admits very large elements (statistical outliers) almost inevitably.

The distribution of the sales of the items typically follows a power-law distribution \cite{Bentley, Fenner, Hisano, Brabazon}, which means that popular items sell far better than niche items.
At the same time, a very large number of niche items can bring a non-negligible percentage of sales,
and this property is called the \textit{long-tail phenomenon} \cite{Anderson}.
By applying the ideas of the critical exponents and universality class from statistical physics \cite{Stanley, Buchanan}, the power-law exponent is interpreted as giving information about market and consumer behavior. 
However, the meaning of the exponent and the process by which it is determined remain unclear.

The aim of this letter is to explain a simple and general mechanism of power-law sales distributions.
We propose a simple stochastic model for an item's selling process.
This model produces a stationary power-law distribution, which corresponds to the distribution of the sales amount.
Moreover, in order to show that the proposed model is comparable with actual process, we carry out analysis of data on U.S. movie income.

\begin{figure}[b!]\centering
\includegraphics[scale=0.9]{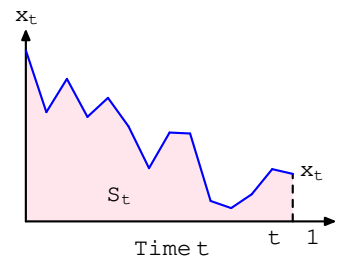}
\caption{
Illustration of the proposed model \eqref{eq1}.
$S_t$ is expressed as the area under the $x_t$ curve.
}
\label{fig1}
\end{figure}

The stochastic model we propose is in the following form:
\begin{subequations}
\label{eq1}
\begin{align}
x_{t+1}&=\mu_t x_t, \label{eq1a}\\
S_{t+1}&=S_t + x_t, \label{eq1b}
\end{align}
\end{subequations}
where $t = 1, 2, \dots$.
This set of equations expresses a selling process of a particular item.
We regard $x_t$ as the sales amount in $t$th period,
and set the initial condition $x_1=\mathrm{constant}$ for simplicity.
We set $S_1=0$, so that $S_t = x_1 + \cdots + x_{t-1}$ represents the total sales amount up to $t-1$.
Figure~\ref{fig1} depicts the illustration of model~\eqref{eq1}.
In the graph of $x_t$ as a function of $t$, $S_t$ is given by the area under the $x_t$ curve.
The stochastic process $\{x_t\}$ given by Eq.~\eqref{eq1a} is known as the Gibrat process \cite{Gibrat}, and $x_t$ approximately follows the log-normal distribution for large~$t$~\cite{Crow}.
Note that Eq.~\eqref{eq1a} implies positive feedback of the sales amount.
That is, if an item sells well in a certain time period, it tends to make a large profit also in the following period.
The growth rate $\mu_t$ is a random number, and for simplicity, we assume that $\mu_t$ is independent and drawn from the same probability distribution for different~$t$.
The accumulated sales amount $S_t$ becomes stochastic by the fluctuation of $\mu_t$.

\begin{figure*}\centering
\includegraphics[scale=0.9]{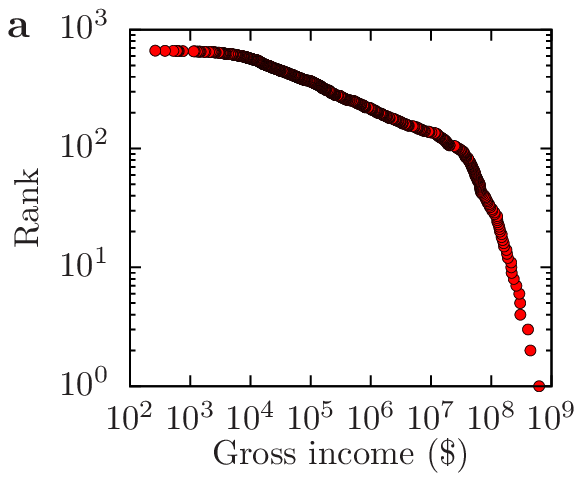}\hspace{3mm}
\includegraphics[scale=0.9]{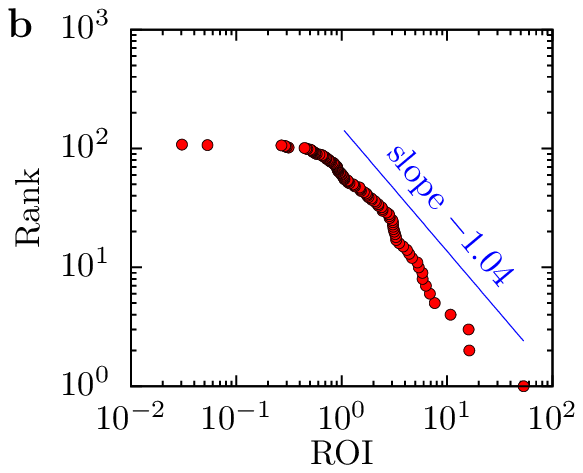}\hspace{3mm}
\includegraphics[scale=0.9]{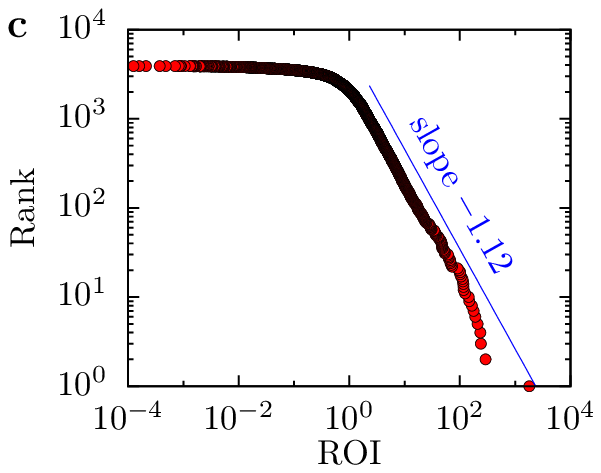}
\caption{
Cumulative distribution of gross income and return on investment (ROI). 
(a) U.S. gross domestic income of 665 movies released in 2012 (from \textit{Box Office Mojo}). 
(b) ROI of 108 movies in 2012. 
(c) ROI of 3906 movies listed in \textit{The Numbers} database. 
The ROI distributions have power-law tails with exponents of (b) $-1.04$ and (c) $-1.12$.
}
\label{fig2}
\end{figure*}

Analysis \cite{Yamamoto} of Eq.~\eqref{eq1} reveals that for large $t$, the cumulative distribution of $S_t$ has a stationary power-law tail $P(S_t>s)\propto s^{-\beta}$, which corresponds to the power-law sales distribution.
The power-law exponent $\beta$ is given by the positive solution of
\begin{equation}
E(\mu_t^\beta)=1, \label{eq2}
\end{equation}
where $E(\cdot)$ stands for the average.
Note that the exponent~$\beta$ is determined only from $\mu_t$.

Now, we check whether the proposed model~\eqref{eq1} captures the essence of actual data.
We analyzed movie income, because sufficient information for our study is available free online.
We used a week as the unit of time $t$ in Eqs.~\eqref{eq1} and \eqref{eq2} in order to smooth out the daily fluctuations, e.g., larger audiences on weekends.
Intuitively, it is plausible to consider that the variable $S_t$ at large $t$ corresponds to gross income, but this perspective has a problem.
We emphasize here that $x_t$ and $S_t$ in Eq.~\eqref{eq1} are random variables which describe the selling process of a particular item (movie title).
We cannot directly compare the empirical distribution of gross income made by gathering a number of different movies and the theoretical distribution of $S_t$ (especially the stationary power law $P(S_t>s)\propto s^{-\beta}$),
because income of different movies are generally not statistically similar to each other.
In fact, the income is associated with the financial scale of the movie, and an expensive movie tends to earn large income~\cite{Pan}.

Here we deduce an appropriate measure for our study, in place of the movie income.
For simplicity, we assume that different movies, namely a pair of random variables $(\mv{x}{i}{t}, \mv{S}{i}{t})$ for the movie $i$, share the same dynamics~\eqref{eq1},
which means that the growth rate $\mu_t$ has the same distribution for different movies.
The scale of the movie $i$, which is denoted by $b^{(i)}$, is included only in the initial value as $\mv{x}{i}{1}=xb^{(i)}$, where $x$ is a constant.
Obviously $\mv{S}{i}{t}$ and $\mv{x}{i}{t}$ depend on the scale $b^{(i)}$,
so they are not statistically similar for different $i$.
On the other hand, $\mv{\tilde{x}}{i}{t}=\mv{x}{i}{t}/b^{(i)}$ and $\mv{\tilde{S}}{i}{t}=\mv{S}{i}{t}/b^{(i)}$ again follows Eq.~\eqref{eq1}, whose initial values $\mv{\tilde{x}}{i}{1}=x$ and $\mv{S}{i}{1}=0$ are constant for all $i$.
Thus, $\mv{\tilde{S}}{i}{t}$, instead of income $\mv{S}{i}{t}$, is statistically similar for different $i$.
In this study, we chose the production budget for $b^{(i)}$;
$\mv{\tilde{S}}{i}{t}$ at large $t$, corresponding to the gross income of the movie $i$ divided by its production budget, is called \textit{return on investment} (ROI).
The ROI is a measure of a movie's success; a movie is in the black if its ROI is greater than 1 and is in the red otherwise.
We expect the empirical distribution of ROI to be comparable with the theoretical stationary distribution of $S_t$.

\begin{figure}[htb!]\centering
\includegraphics[scale=0.9]{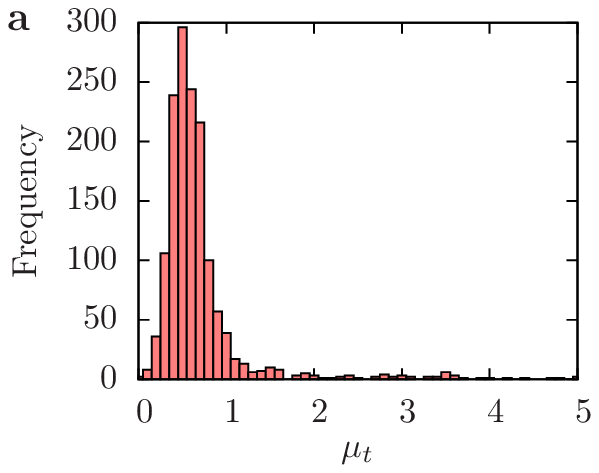}\vspace{3mm}
\includegraphics[scale=0.9]{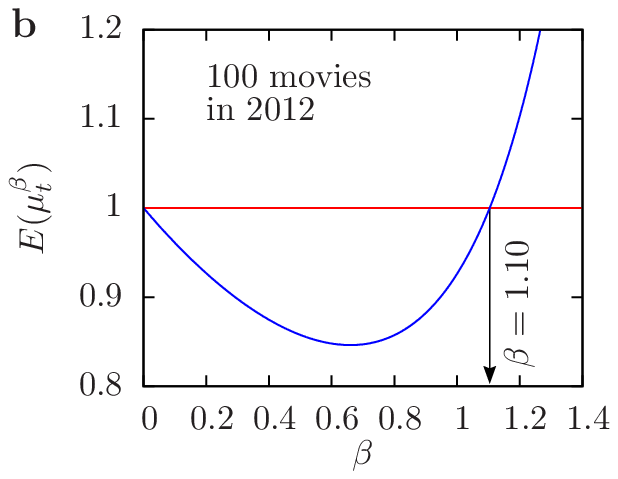}\vspace{3mm}
\includegraphics[scale=0.9]{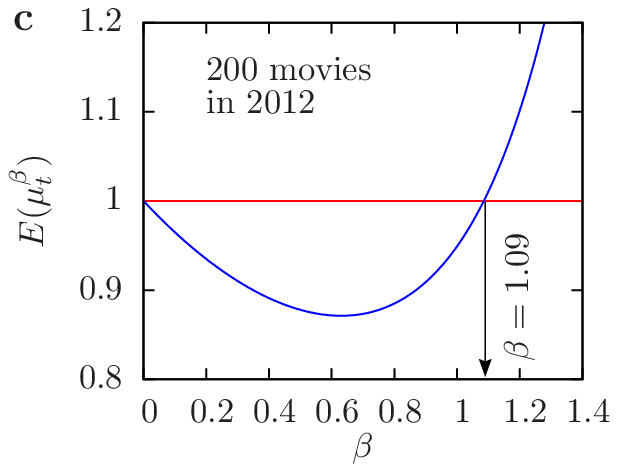}\vspace{3mm}
\includegraphics[scale=0.9]{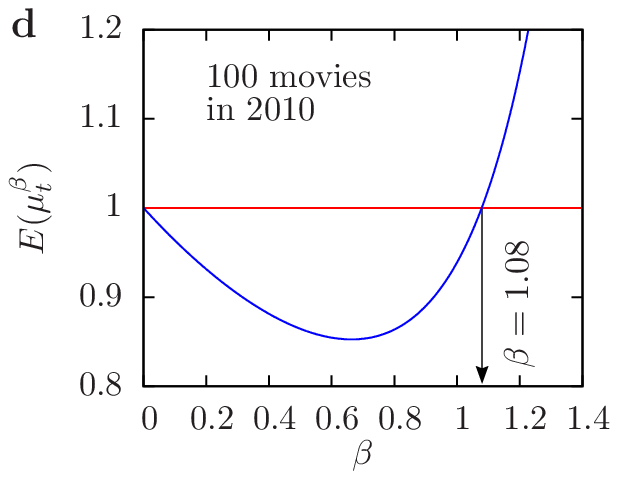}
\caption{
Estimate of power-law exponent $\beta$. 
(a) Frequency histogram of the ratio $\mu_t$ of weekly income between two adjoining weeks compiled from top-100 movies in 2012. 
(b) To determine $\beta$, we calculated $E(\mu_t^\beta)$ as a function of $\beta$, by using top-100 movies in 2012. 
The positive solution of $E(\mu_t^\beta)=1$ is $\beta=1.10$. 
(The trivial solution $\beta=0$ always exists, but it is irrelevant to the power-law exponent.) 
(c) $\beta=1.09$ is obtained from top-200 movies in 2012.
(d) $\beta=1.08$ is obtained from top-100 movies in 2010.
}
\label{fig3}
\end{figure}

Figure~\ref{fig2}a shows the cumulative distribution of U.S. gross domestic income for 665 movies released in 2012.
This set of data is collected from the free online database \textit{Box Office Mojo}\cite{BoxOfficeMojo}.
This distribution does not possess a clear power-law tail.
There are various opinions about the distribution of gross income;
some reports claim that the gross income follows a power law \cite{Sornette, Sinha}, whereas others are not \cite{Pan, Walls}.
In contrast, Fig.~\ref{fig2}b shows the distribution of ROI of 108 movies that came out in 2012,
which exhibits heavy-tailed behavior. 
We could compute the ROI of only 108 movies because the production budgets of the other movies were not in \textit{Box Office Mojo} database. 
By using the method of maximum likelihood~\cite{Clauset},
the power-law exponent of the tail, corresponding to ROI larger than $10^0$, is calculated as $\beta=1.04$,
and this power law is shown as the straight line in Fig.~\ref{fig2}b.
The standard error on $\beta$ is $\sigma=0.13$;
$\sigma$ becomes large because only 68 movies in Fig.~\ref{fig2}b achieve ROI more than unity.
To obtain more samples, we used another database, \textit{The Numbers}~\cite{TheNumbers}, which provides a more complete list of production budgets and allowed us to compute the ROI of 3906 movies from 1915 to April 2014. 
The ROI distribution generated by this list, shown in Fig.~\ref{fig2}c, also has a power-law tail.
The estimated power-law exponent is $\beta=1.12\pm0.03$ in the range where ROI is more than $2$, indicated by the straight line.
From these results, ROI approximately follows Zipf's law \cite{Zipf}.
The following is the discussion that the distribution of $S_t$ from Eq.~\eqref{eq1} is consistent with the actual distribution of ROI.

Let us check whether Eq.~\eqref{eq2} gives a satisfactory estimate of $\beta$, assuming that actual movie income follows model \eqref{eq1}.
The variable $\mu_t$ is calculated as the ratio of the income on week $t+1$ to that on $t$ (i.e., $\mu_t=x_{t+1}/x_t$). 
In Fig.~\ref{fig3}a, we show the histogram of $\mu_t$ built from the weekly records of top-100 movies released in 2012 obtained from \textit{Box Office Mojo} database. 
By using the set of $\mu_t$'s, we calculate $E(\mu_t^\beta)$ as a function of $\beta$ (Fig.~\ref{fig3}b), and find that $\beta=1.10$ is the positive solution of $E(\mu_t^\beta)=1$. 
This solution is close to the exponent $\beta=1.04$ and $1.12$ obtained from the ROI data presented in Fig.~\ref{fig2}. 
Similarly, we find $\beta=1.09$ and $1.08$ from the top-200 movies in 2012 (Fig.~\ref{fig3}c) and the top-100 movies in 2010 (Fig.~\ref{fig3}d), respectively. 
That is, the estimate of $\beta$ does not depend strongly on the rank or the released year. 
Based on model~\eqref{eq1}, we obtain a good estimate of $\beta$, and this result supports the validity of the model.

\begin{figure}[tb!]\centering
\includegraphics[scale=0.9]{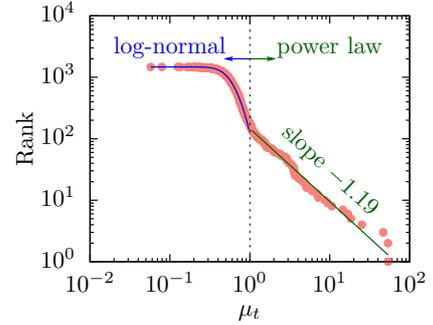}
\caption{
Cumulative distribution of $\mu_t$'s built from the record of top-100 movies in 2010.
A log-normal distribution is suitable in $\mu_t<1$, and power-law decay appears in $\mu_t>1$.
}
\label{fig4}
\end{figure}

In Fig.~\ref{fig3}a, $\mu_t$ mostly falls between 0 and 1,
but there exist a few large values.
To see this in more detail,
we give further statistical analysis for the distribution of $\mu_t$.
The overall cumulative distribution of $\mu_t$ is shown in Fig.~\ref{fig4}.
Clearly, behavior critically changes at $\mu_t=1$;
the distribution of $\mu_t$ approximately follows a log-normal distribution in $\mu_t<1$, while the distribution roughly exhibits power-law decay, having exponent $-1.19$, in $\mu_t>1$.
The discontinuous crossover at $\mu_t=1$ perhaps causes a great disparity between hit movies and poor movies.
(Power-law exponent $-1.19$ is not directly related to the exponent~$\beta$ of the ROI distribution.)

\begin{figure}\centering
\includegraphics[scale=0.9]{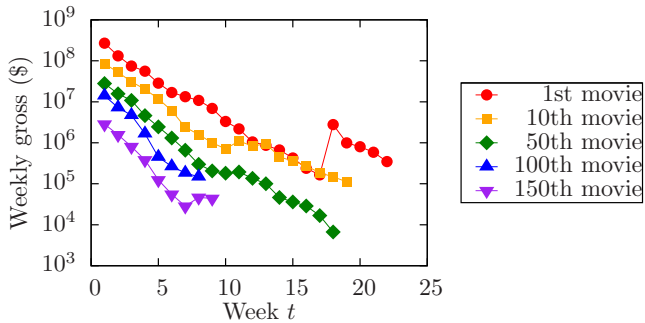}
\caption{
Exponential decay of weekly gross income. 
The weekly income of the 1st, 10th, 50th, 100th, and 150th highest-grossing movies in 2012 (\textit{Marvel's the Avengers}, \textit{Madagascar 3}, \textit{Chronicle}, \textit{The Five-Year Engagement}, and \textit{Casa de Mi Padre}, respectively) decrease exponentially with a common decay rate. 
The curve for the 1st movie has the jump from week 17 to week 18, which is mainly because theaters showing this movie increased sharply in week 18. 
The curves for the 100th and 150th movies end before week 10 because weekly records are inclined to be incomplete in lower-ranked movies.
}
\label{fig5}
\end{figure}

Next, we discuss a statistics of an individual movie. 
By taking the logarithm of Eq.~\eqref{eq1a}, we obtain
\begin{equation*}
\ln x_{t+1} = \ln x_t + \ln \mu_t.
\end{equation*}
This equation means that $x_t$ is a random walk on a logarithmic scale and its average displacement per step is $E(\ln\mu_t)$. 
Thus, on a linear scale, $x_t$ typically decreases exponentially according to $x_t\propto\exp(-\gamma t)$, where $\gamma=-E(\ln\mu_t)$ is the decay rate. 
This theoretical outcome is qualitatively correct for actual movies. 
Figure~\ref{fig5} shows the weekly income of the 1st, 10th, 50th, 100th, and 150th highest grossing movies in the U.S. released in 2012 and confirms their exponential decay.
Their decay rates are respectively
$\gamma_1=0.453$, $\gamma_{10}=0.561$, $\gamma_{50}=0.607$, $\gamma_{100}=0.716$, and $\gamma_{150}=0.598$.
(The unit of $\gamma$ is $\mbox{week}^{-1}$.)
On the other hand, by using the distribution of $\mu_t$ shown in Figs.~\ref{fig3}a and \ref{fig4}, we obtain $-E(\ln\mu_t)\allowbreak=0.416\mbox{week}^{-1}$. 
Mathematically, $\gamma>0$ (equivalently $E(\ln\mu_t)<0$) is a necessary condition for $S_t$ to have a stationary power-law distribution \cite{Yamamoto}.
A previous study \cite{Pan} reported the exponential decay of daily movie income, but without proposing a simple mechanism like Eq.~\eqref{eq1}. 
Furthermore, the weekly income per theater of a movie shows a power-law decay in time \cite{Pan},
but this result cannot be obtained by the model in this letter.

\begin{figure}[t!]\centering
\includegraphics[scale=0.9]{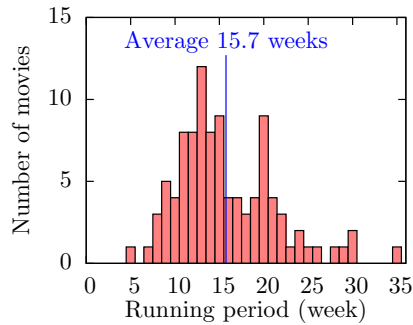}
\caption{
Frequency histogram of the running weeks of top-100 movies in 2012.
The average is $15.7$ weeks.
}
\label{fig6}
\end{figure}

A movie has a running period for which it is shown in theaters.
Yet, we have compared the actual ROI distribution and the stationary distribution of $S_t$,
without considering the effect of finite running period in the model.
We can properly exclude it, based on the discussion as follows.
As shown in Fig.~\ref{fig5}, the weekly income decreases approximately exponentially.
Hence, income in first several weeks accounts for a large percentage of the gross income.
More accurately, if a movie has $T$ running weeks and its weekly income follows exact exponential decay $x_t=x\exp(-\gamma(t-1))$,
the gross income is given by
$$
S_{T+1}=\sum_{t=1}^{T} x_t = \frac{1-e^{-\gamma T}}{1-e^{-\gamma}}x.
$$
The gross income of this movie if it has infinite running weeks is
$S_\infty = x/(1-\exp(-\gamma))$,
which corresponds to the theoretical stationary value of the gross income.
The ratio of $S_{T+1}$ to $S_\infty$ is 
$$
\frac{S_{T+1}}{S_\infty}=1-e^{-\gamma T}.
$$
A typical value of the decay rate is $\gamma=0.5\mbox{week}^{-1}$,
and the average running period of top-100 movies in 2012 is $T=15.7$ weeks (see Fig.~\ref{fig6} for reference).
Using these values, we can estimate as $S_{T+1}/S_\infty\approx0.9996$.
Therefore, $S_T$ is essentially the same as $S_\infty$.
In other words, income $S_t$ becomes almost stationary within typical running weeks.
This is why we can ignore the effect of finite running weeks.

We conclude that the power-law behavior of the ROI of movies is adequately simulated by Eq.~\eqref{eq1}. 
Meanwhile, we need to be careful with data bias.
The list of the production budgets \cite{TheNumbers} is incomplete for movies having low production budgets.
Moreover, the listed budgets are rough estimates, and tend to be less reliable in comparison to the record of gross income.
We need to study other sales distributions, so as to ascertain that our theoretical model~\eqref{eq1} is applicable to actual data.

We consider Eq.~\eqref{eq1} to be the minimal model for the power-law sales distribution.
It has a very simple form. 
It does not directly consider the effects of consumers' preference, advertisement, and word of mouse~\cite{Ishii}; these factors are condensed into the random variable $\mu_t$.
Recall that the aim of this letter is to give a simple mechanism for the power-law behavior.
We do not intend a detailed and faithful description of a real phenomenon.
Elaborate analysis of a movie market deals with miscellaneous statistical data~\cite{Vany}.
Although it is difficult to predict from the proposed model whether a specific movie succeeds, it can be useful to understand the overall market situation. 
Moreover, by virtue of the simplicity, we expect the proposed model to be applicable to heavy-tailed behavior other than movie income. 
In particular, our model can lead to more efficient simulations of social and economic phenomena, especially in the domain of market research.
The combination of a time-dependent quantity $x_t$ and its summation $S_t=x_1+\dots+x_{t-1}$ concisely captures history dependence. 
Therefore, we anticipate that the proposed model becomes a theoretical basis for describing a variety of power-law behavior in social phenomena. 

\section*{Acknowledgments}
The author is grateful to Professor Yoshihiro Yamazaki for constructive comments. 
The present work was supported by a Grant-in-Aid for Young Scientists (B) (25870743) from the Ministry of Education, Culture, Sports, Science, and Technology (MEXT) of Japan.

\balance
\end{document}